%% file: main.tex
\documentclass[conference]{IEEEtran}

\usepackage{import}
\usepackage[para]{threeparttable}
\usepackage{booktabs}
\usepackage{tabularx}
\usepackage{tikz}
\usepackage{pifont}
\usepackage{subcaption}
\usepackage[binary-units]{siunitx}
\usepackage{textcomp}

\usepackage{array}
\newcolumntype{P}[1]{>{\centering\arraybackslash}p{#1}}
\newcolumntype{M}[1]{>{\centering\arraybackslash}m{#1}}

\import{./}{styles.tex}

\usetikzlibrary{arrows,shapes}

\newcounter{observationcounter}

\begin{document}

\IEEEoverridecommandlockouts
\IEEEpubid{\begin{minipage}[t]{\textwidth}\ \\[10pt]
        \centering\normalsize{979-8-3503-4099-0/23/\$31.00  \copyright 2023 IEEE}
\end{minipage}} 

\title{When Memory Mappings Attack: On the (Mis)use of the ARM Cortex-M FPB Unit}

\author{
\IEEEauthorblockN{Haoqi Shan}
\IEEEauthorblockA{CertiK \\
haoqi.shan@certik.com}
\and
\IEEEauthorblockN{Dean Sullivan}
\IEEEauthorblockA{University of New Hampshire \\
dean.sullivan@unh.edu}
\and
\IEEEauthorblockN{Orlando Arias}
\IEEEauthorblockA{University of Massachusetts, Lowell \\
orlando\_arias@uml.edu}

}

\maketitle

\begin{abstract}
\import{./abstract/}{abstract.tex}
\end{abstract}

\begin{IEEEkeywords}
Design-for-Debug, Embedded Security, ARM Cortex-M
\end{IEEEkeywords}

\section{Introduction}\label{sec:introduction}
\import{./introduction/}{introduction_v2.tex}

\section{Background}\label{sec:background}
\import{./background/}{background_v2.tex}

\section{Delving into the FPB}\label{sec:design}
\import{./design/}{design_v2.tex}

\section{Deploying the Attacks}\label{sec:evaluation}
\import{./evaluation/}{evaluation.tex}

\section{Security Arms Race on the Embedded Field}\label{sec:related}
\import{./related/}{related.tex}

\section{Conclusion}\label{sec:conclusion}
\import{./conclusion/}{conclusion.tex}

\bibliographystyle{IEEEtran}
\bibliography{bib/references}

\end{document}

%% file: styles.tex
\usepackage{xcolor}
\usepackage{textcomp}
\usepackage{listings}

\usepackage{armasm}

\definecolor{comment}{rgb}{0,.6,0}
\definecolor{keyword}{rgb}{.63,0,.42}
\definecolor{kw2}{rgb}{.50,.50,.15}
\definecolor{kw3}{rgb}{.42,.42,.63}
\definecolor{string}{rgb}{1,0,0}

\lstdefinestyle{c}{
	language=C,
    commentstyle=\color{comment},
	stringstyle=\color{string},
	keywordstyle=\bfseries\color{keyword},
	basicstyle=\footnotesize\ttfamily,
	numbers=left,
	numberstyle=\tiny,
	numbersep=5pt,
	frame=lines,
	breaklines=true,
	prebreak=\raisebox{0ex}[0ex][0ex]{\ensuremath{\hookleftarrow}},
	showstringspaces=false,
	upquote=true,
	tabsize=2,
    xleftmargin=2em,
    xrightmargin=2em,
}

\lstdefinestyle{arma}{
	language=[armasm]Assembler,
    commentstyle=\color{comment},
	stringstyle=\color{string},
	keywordstyle=[1]\bfseries\color{keyword},
	keywordstyle=[2]\bfseries\color{kw2},
	keywordstyle=[3]\bfseries\color{kw3},
	basicstyle=\footnotesize\ttfamily,
	numbers=left,
	numberstyle=\tiny,
	numbersep=5pt,
	frame=lines,
	breaklines=true,
	prebreak=\raisebox{0ex}[0ex][0ex]{\ensuremath{\hookleftarrow}},
	showstringspaces=false,
	upquote=true,
	tabsize=2,
    xleftmargin=2em,
    xrightmargin=2em,
}

\makeatletter
\pgfdeclareshape{obsframe}{
  \inheritsavedanchors[from=rectangle]
  \inheritanchorborder[from=rectangle]
  \inheritanchor[from=rectangle]{center}
  \inheritanchor[from=rectangle]{base}
  \inheritanchor[from=rectangle]{north}
  \inheritanchor[from=rectangle]{north east}
  \inheritanchor[from=rectangle]{east}
  \inheritanchor[from=rectangle]{south east}
  \inheritanchor[from=rectangle]{south}
  \inheritanchor[from=rectangle]{south west}
  \inheritanchor[from=rectangle]{west}
  \inheritanchor[from=rectangle]{north west}
  \backgroundpath{
    \southwest \pgf@xa=\pgf@x \pgf@ya=\pgf@y
    \northeast \pgf@xb=\pgf@x \pgf@yb=\pgf@y
    \pgfpathmoveto{\pgfpoint{\pgf@xa}{\pgf@ya}}
    \pgfpathlineto{\pgfpoint{\pgf@xb}{\pgf@ya}}
    \pgfpathmoveto{\pgfpoint{\pgf@xa}{\pgf@yb}}
    \pgfpathlineto{\pgfpoint{\pgf@xb}{\pgf@yb}}
 }
}
\makeatother

%% file: abstract/abstract.tex
In recent years we have seen an explosion in the usage of low-cost, low-power
microcontrollers (MCUs) in embedded devices around us due to the popularity of
Internet of Things (IoT) devices. Although this is good from an economics
perspective, it has also been detrimental for security as microcontroller-based
systems are now a viable attack target. In response, researchers have developed
various protection mechanisms dedicated to improve security in these
resource-constrained embedded systems.

We demonstrate in this paper these defenses fall short when we leverage benign
memory mapped design-for-debug (DfD) structures added by MCU vendors in their
products. In particular, we utilize the Flash Patch and Breakpoint (FPB) unit
present in the ARM Cortex-M family to build new attack primitives which can be
used to bypass common defenses for embedded devices.
Our work serves as a warning and a
call in balancing security and debug structures in modern microcontrollers.

%% file: introduction/introduction_v2.tex
We have seen a surge in usage of deeply embedded systems accompanied with the advent of low-cost, low-power microcontrollers (MCUs). These devices are
designed to meet long deployment times in unattended environments. Commonly,
these devices run software in a freestanding (bare-metal) environment with
programmers being responsible for direct management of all hardware resources.

Hardware vendors typically augment their platforms with built-in debug and profiling facilities to alleviate developer burden.
These aid in
facilitating the creation of software development tools by making this
functionality available to the on-chip scan chain and accessible to the
software itself. For example, the Flash Patch and Breakpoint (FPB) Unit and
the Data Watchpoint and Trace (DWT) Unit found in the ARMv6-M, ARMv7-M, and
ARMv8-M architectures provide an universally available Designed-for-Debug (DfD)
infrastructure in Cortex-M devices. These peripherals can easily be configured
to set up debug events and provide performance monitoring capabilities in
embedded devices.

Even with a plethora of debugging tools software bugs still abound leading to
exploitable vulnerabilities such as improper data validation
\cite{rosenberg2014reflections, CVE-2015-4421, CVE-2015-4422}, memory
corruption \cite{CVE-2015-6639, CVE-2018-16522, CVE-2018-16526}, and improper
permission checks \cite{CVE-2017-13209}. These vulnerabilities, compounded by
the fact that embedded software rarely uses separation of privileges and task
isolation, have resulted in sophisticated and disastrous attacks against
embedded devices \cite{cui2013firmware, fur2016exploiting, garcia2017hey,
ronen2017iot}. As a result, multiple defenses have been introduced to prevent
or mitigate attacks against embedded devices \cite{nyman2017cfi,
koeberl2014trustlite, brasser2015tytan, clements2017protecting,
kim2018securing}. However, we note that benign DfD interfaces such as the FPB
unit remain accessible to software through the microcontroller's memory map
well after the software development process ends.

In this paper, we demonstrate how
we can regain arbitrary control of bare-metal devices in lieu of the latest
defenses through the use of these DfD interfaces without the need of using a
debug probe.
In particular, we show how a carefully designed FPB configuration is capable of bypassing memory protection schemes in microcontrollers resulting on data leakage.
We demonstrate how our
crafted primitives work in spite of the threat model assumed by the defenses,
thereby proving the inherent incompatibility of software-accessible DfD
interfaces and security mechanisms in embedded devices.

In summary, the contributions of this paper are:
\begin{itemize}
	\itemsep0em
	\item A new set of exploitation primitives that leverage memory mapped DfD
interfaces in Cortex-M microcontrollers. We show how we can achieve persistence
without the need of a debug probe.
	\item Evaluation of the attack primitives and a demonstration of its
applicability bypassing memory protection and leaking data from an embedded device.
\end{itemize}


The remainder of this paper is structured as follows: Section
\ref{sec:background} provides high level overview of the Cortex-M cores,
Section \ref{sec:design} provides a look at the functionality of the
FPB, Section \ref{sec:evaluation} discusses how we utilize the FPB
functionality to deploy different attacks against defense mechanisms.
We discuss 
related
works in Section \ref{sec:related}, and draw concluding remarks in
Section \ref{sec:conclusion}.

%% file: background/background_v2.tex
\subsection{The Cortex-M Processor Family}
The Cortex-M family of processor cores are among the most popular cores used in
embedded devices. Designed and licensed by ARM Limited, the IP has been shipped
in tens of billions of devices \cite{arm-cortex-m-wiki}. Compared to the
Cortex-R and Cortex-A counterparts, Cortex-M cores offer reduced functionality
to achieve the low power requirements of embedded devices.

Cortex-M processors use a flat memory architecture, with memory mapped
peripherals and related functions. The core has two execution modes:
\emph{thread} and \emph{handler}. Application code is run in thread mode, which
can be \emph{privileged} or \emph{non-privileged}. Handler mode is used to
handle exceptions, such as interrupts. Execution in handler mode is always
privileged.
Cortex-M interrupts are vectored, with the region containing the addresses to
the service routines being called the \emph{vector table}. The vector table is
always readable regardless of MPU permissions. Upon starting, the core
initializes \texttt{pc} and \texttt{msp} with the values in their corresponding
entries in the vector table, switches to thread mode and starts privileged
execution.

\subsection{Debug Modes in Cortex-M Microcontrollers}
ARM defines debug events to be those triggered for debug reasons. The actions
taken by a CPU due to a debug event depend on the state of the
\texttt{C\_DEBUGEN} bit in the Debug Handling Control and Status Register,
\texttt{DHCSR}. Although the aforementioned register is memory mapped,
\texttt{C\_DEBUGEN} can only be written to by a debug probe. Setting this bit
to 1 enables \emph{halting debug} mode. Under this condition, the CPU halts
whenever a debug event is triggered. The bit is reset to 0 whenever a debug
probe is disconnected. Under this mode, a debug event normally causes a
\texttt{DebugMonitor} exception.

\subsection{Flash Patch and Breakpoint Unit}\label{sec:fpb_background}
The Flash Patch and Breakpoint (FPB) unit is a hardware module which is
intended for debug and on the fly patching of firmware \cite{armv6m-arm,
armv7m-arm, armv8m-arm}. Although an optional component, we have yet to find a
single Cortex-M device that does not provide this feature. At its most basic
level the FPB is used to add hardware breakpoints during a debug session.  A
full FPB implementation provides the ability to replace instructions and data
fetches performed by the CPU as a way to provide software fixes to otherwise
immutable areas of memory, such as ROMs. We provide more details on the FPB in
Section \ref{sec:design}.


%% file: design/design_v2.tex
In this section we go into detail on the operation of the Flash Patch and
Breakpoint (FPB) Unit and how it can be configured through software.
At a high level, the FPB can remap sections of program memory to regions of
RAM, as well as set breakpoints on code memory. ARM designed this unit for
debugging purposes, and as a way to provide code or data patches to an
application that requires field updates to ROM code. Currently, two versions
of the FPB are available for licensing by microcontroller vendors.
System-on-chip vendors can disable the remap functionality provided by the FPB
at design time. The FPB can be activated and used regardless of the CPU's debug
state.

\subsection{The FPB Unit's Functionality}
A full FPB implementation provides the means to transparently ``patch'' memory
addresses in the lower \SI{512}{\mebi\byte} range of the address space of a
Cortex-M. More properly, a full FPB implementation can transparently override
instruction and data fetches done by the core using data contained in an area
of SRAM called the \emph{remap table}. The FPB can be configured directly by a
debugger or software running on the device using the registers listed in Table
\ref{tb:fpb_register_set}.

\begin{table}[h]
    \centering
    \caption{FPB Register Set. These registers are memory mapped starting at
    address \texttt{0xe0002000}. The nomenclature chosen for the register names
    adheres to the convention used in ARM Architecture
    Reference Manuals \cite{armv7m-arm,armv8m-arm}.}
    \label{tb:fpb_register_set}
    \begin{tabularx}{\columnwidth}{lX}
        \toprule
        \textbf{Register} & \textbf{Function} \\
        \midrule
        \texttt{FP\_CTRL} & Provides FPB implementation information, and the
        global enable for FPB unit. \\
        \texttt{FP\_REMAP} & Indicates whether the implementation supports flash
        patch remapping, and holds the SRAM address for remap. \\
        \texttt{FP\_COMP}\textit{n} & Holds the address for comparison. For
        instruction address comparators, these registers can remap to an
        instruction in an address in SRAM or define a breakpoint. For address
        comparators, this register defines an address to remap to SRAM.\\
        \bottomrule
    \end{tabularx}
\end{table}

\begin{figure*}[h]
	\centering
	\begin{subfigure}[b]{\columnwidth}
		\centering
		\resizebox{\columnwidth}{!}{
			\import{./figs/}{fpb_sample_setup_fetch_insn.tex}
		}
		\caption{The CPU obtains the load instruction in RAM from the
		FPB instead of the one in flash when executing from address
		\texttt{0x08090030}. The resulting operation is a dereference
		from a different address.}
		\label{fig:fpb_setup_demo_insn}
	\end{subfigure}
	\hfill
	\begin{subfigure}[b]{\columnwidth}
		\centering
		\resizebox{\columnwidth}{!}{
			\import{./figs/}{fpb_sample_setup_fetch_data.tex}
		}
		\caption{The CPU fetches the value to load into \texttt{r0} in
		RAM from the FPB rather than the expected \texttt{0x08090034}
		address. This causes a different literal to be stored in
		\texttt{r0}.}
		\label{fig:fpb_setup_demo_data}
	\end{subfigure}
	\caption{Setting up the FPB to intercept (a) instruction fetches, and
	(b) data fetches.}
\end{figure*}
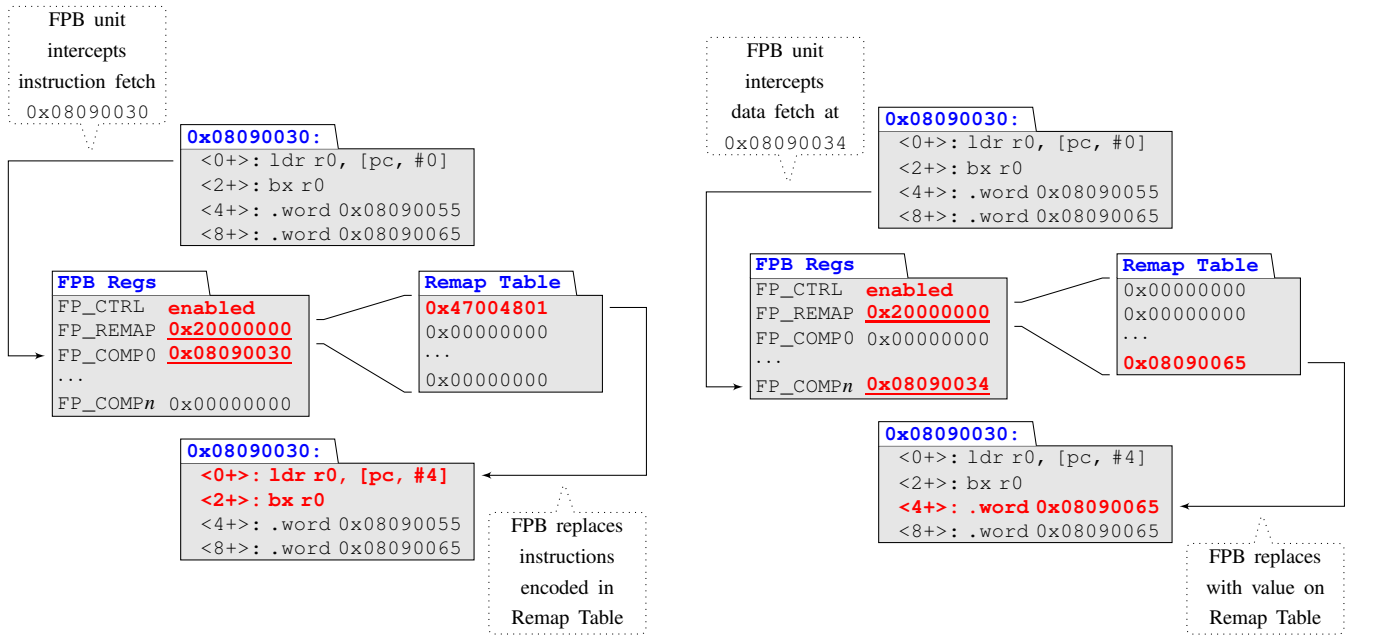

The \texttt{FP\_CTRL} register provides information as to the number of
comparator registers available in the implementation, as well as the functions
of these registers. The \texttt{FP\_COMP}\textit{n} registers provide the means
of identifying the addresses that will be affected by the FPB. If present, the
\texttt{FP\_REMAP} register can point to a location in SRAM containing the
remap table. The remap table contains the data or instructions that will be
seen by the CPU whenever the latter accesses the addresses signaled to by the
FPB comparator registers. Since all these registers are memory mapped, software
running on the platform can change them. We make the observation that
\emph{software stored in a read only memory (e.g. a mask ROM) can still be made
self modifying by utilizing the FPB}.

Another characteristic of the FPB is that it bypasses the MPU's configuration.
When the FPB accesses the remap table, it ignores any MPU settings for that
particular region of SRAM. For example, if code is executing at non-privileged
level, and the remap table is located in a region only readable by privileged
software the ensuing memory access by the FPB will not raise the
\texttt{MemManage} exception. Similarly, the remap table could be in a region
of memory flagged as non-executable and redirection of instruction fetches
performed by the FPB not trigger the \texttt{MemManage} exception. Thus,
\emph{application software can read data and execute code for an arbitrary
location of SRAM regardless of MPU settings for that area by means of the remap
functionality}.

Lastly, outside halting debug mode (\texttt{DHCSR.C\_DEBUGEN} is unset and
no JTAG probe is active), execution from an address flagged by the FPB as
containing a breakpoint raises a \texttt{DebugMonitor} exception provided that
the exception is enabled. If the group priority of the \texttt{DebugMonitor}
exception is equal or less than the current execution priority, the exception
will escalate into a \texttt{HardFault}. This has the implication that \emph{an
application can insert breakpoints in its own code by using the FPB and proceed
to monitor its own execution through the \texttt{DebugMonitor} exception}.

\subsection{Towards the Attack Primitives}

We show an example FPB setup in Figure \ref{fig:fpb_setup_demo_insn}. Here, we
wish to modify the instruction stream executed by the processor by changing the
location from which the \lstinline[style=arma]{ldr} (load register) instruction
will act on. To perform this, we enable the code comparator \texttt{FP\_COMP0}
and have it target the address of the load instruction. We also set up a remap
table in a viable location in RAM. The first entry of the remap table contains
the encoding of the instruction we wish to execute, other entries in the table
can be set to anything else. To better illustrate this example, we set all
other entries in the remap table to 0. The instruction decodes to a load
register, but from a different offset. This ultimately results in the value
loaded into \lstinline[style=arma]{r0} being different from what was originally
intended.

Figure \ref{fig:fpb_setup_demo_data} shows the same code example but with a
modified FPB configuration. The new configuration utilizes a data literal
comparator register to change the value loaded into \lstinline[style=arma]{r0}
from the load instruction's offset. This results in the indirect branch
instruction targeting a different address in program space.

We are now in position to generalize the FPB setup for our attack primitives.
Our goal is to achieve arbitrary code execution or leak data in the device in
spite of protections offered. To achieve this, we alter the functionality
of the application by writing to the FPB registers. As such, we:

\begin{enumerate}
	\itemsep0em
	\item Configure the FPB registers by setting the comparator registers
	to the desired area to be patched.

	\item Inject the remap table containing the instructions and literals we
	wish to execute or read, then enable the FPB.

	\item Let execution proceed. When the targeted address/data declared in the FPB
    comparator registers is reached, the FPB will mask the instruction stream
    with the one in the remap table. When data literal accesses are made, the
    FPB will provide the values stored in the remap table.
\end{enumerate}

\subsection{Baseline Test and Example}
As a baseline test we configured the FPB to redirect control-flow to an
otherwise unused function we placed in flash memory.
We set an FPB code comparator register to capture an instruction fetch at the
first address of the reset vector handler. At the corresponding entry in the
remap table, we encode a branch instruction to our payload function.
Once the FPB is set up, we trigger a software reset of the microcontroller by
writing to the \texttt{AIRCR} register. The microcontroller loads the address
of the reset vector handler from the vector table and starts executing from
there. Since the FPB is active and a comparator register points to that
address, the CPU sees the instruction we stored in the remap table in
RAM and executes it. Our target function proceeds to execute.

%% file: figs/fpb_sample_setup_fetch_insn.tex
\begin{tikzpicture}
	\tikzstyle{insnlbl} = [
			text width=4cm,
			inner sep=0pt,
			xshift=2pt,
			font=\footnotesize,
			anchor=west,
		];
	
	\tikzstyle{datalbl} = [
			style=insnlbl,
			font=\footnotesize\ttfamily,
		];

	\def\insnlist{
			\textbf{0x08090030:}/blue,
			~~<0+>: ldr r0{,} [pc{,} \#0]/black,
			~~<2+>: bx r0/black,
			~~<4+>: .word 0x08090055/black,
			~~<8+>: .word 0x08090065/black%
		};

	\def\insnlistpatch{
			\textbf{0x08090030:}/blue,
			~~\textbf{<0+>: ldr r0{,} [pc{,} \#4]}/red,
			~~\textbf{<2+>: bx r0}/red,
			~~<4+>: .word 0x08090055/black,
			~~<8+>: .word 0x08090065/black%
		};
	
	\def\remaplist{
			\textbf{0x47004801}/red,
			0x00000000/black,
			$\cdots$/black,
			0x00000000/black%
		};
	
	\def\reglist{
			\texttt{FP\_CTRL}/\textbf{enabled}/red,
			\texttt{FP\_REMAP}/\textbf{\underline{0x20000000}}/red,
			\texttt{FP\_COMP0}/\textbf{\underline{0x08090030}}/red,
			$\cdots$/~/black,
			\texttt{FP\_COMP}\textit{n}/0x00000000/black%
		};

	\draw[color=black, fill=gray!20, thin] (1.75, -4.785) rectangle (5.75, -6.105);
	\draw[color=black, fill=gray!20, thin] (1.75, -.495) rectangle (5.75, -1.815);

	\draw[color=black, fill=gray!20, thin] (0, -2.495) rectangle (3.5, -4.145);
	\draw[color=black, fill=gray!20, thin] (5, -2.495) rectangle (7.5, -3.815);

	\draw[color=black, fill=white!10, thin] (1.75, -4.785)
			-- ++(0, .33)
			-- ++(2.1, 0)
			-- ++(0.05, -.33);
	\draw[color=black, fill=white!10, thin] (1.75, -.495)
			-- ++(0, .33)
			-- ++(2.1, 0)
			-- ++(0.05, -.33);
	\draw[color=black, fill=white!10, thin] (0, -2.495)
			-- ++(0, .33)
			-- ++(2.1, 0)
			-- ++(0.05, -.33);
	\draw[color=black, fill=white!10, thin] (5, -2.495)
			-- ++(0, .33)
			-- ++(2.1, 0)
			-- ++(0.05, -.33);

	\node[xshift=2pt, inner sep=0pt, color=blue,anchor=west,font=\footnotesize]
			at(0, -2.33) {\texttt{\textbf{FPB Regs}}};
	\node[xshift=2pt, inner sep=0pt, color=blue,anchor=west,font=\footnotesize]
			at(5, -2.33) {\texttt{\textbf{Remap Table}}};

	\foreach \insn\colour [count=\i] in \insnlist  {
		\node[style=datalbl,color=\colour] at(1.75, -0.33*\i) {\insn};
	}

	\foreach \insn\colour [count=\i] in \insnlistpatch  {
		\node[style=datalbl,color=\colour] at(1.75, -4.29 -0.33*\i) {\insn};
	}

	\foreach \insn\colour [count=\i] in \remaplist  {
		\node[style=datalbl,color=\colour,text width=2cm] at(5, -2.33-0.33*\i) {\insn};
	}

	\foreach \reg\val\colour [count=\i] in \reglist {
		\node[style=insnlbl] at(0, -2.33-0.33*\i) {\reg};
		\node[style=datalbl,color=\colour] at(1.5, -2.33-0.33*\i) {\val};
	}

	\draw[-latex'] (1.65, -0.66) -- ++(-2.25, 0)
		node[anchor=south west,inner sep=2pt, text width=2cm, draw, rectangle
		callout, rounded corners=2pt, callout absolute pointer={(0.525, -.56)},
		fill=white, shift={(0, .5)}, dotted, text centered]
			{\footnotesize FPB unit intercepts instruction fetch \texttt{0x08090030}}
		|- (-0.1, -2.66-0.33*2);

	\draw (3.6, -2.825) -- ++ (.1, 0) -- (4.8, -2.495) -- ++ (.1, 0);
	\draw (3.6, -3.155) -- ++ (.1, 0) -- (4.8, -3.815) -- ++ (.1, 0);

	\draw[-latex'] (7.6, -2.66) -- ++ (0.5, 0) |-
		node[draw, dotted, rectangle callout, rounded corners=2pt, fill=white, anchor=north east, inner sep=2pt,
		text width=2cm, shift={(0, -.5)},
		callout absolute pointer={(6.975,-5.05)}, text centered]
		{\footnotesize FPB replaces instructions encoded in Remap Table} (5.85, -4.95);
	
\end{tikzpicture}

%% file: figs/fpb_sample_setup_fetch_data.tex
\begin{tikzpicture}
	\tikzstyle{insnlbl} = [
			text width=4cm,
			inner sep=0pt,
			xshift=2pt,
			font=\footnotesize,
			anchor=west,
		];
	
	\tikzstyle{datalbl} = [
			style=insnlbl,
			font=\footnotesize\ttfamily,
		];

	\def\insnlist{
			\textbf{0x08090030:}/blue,
			~~<0+>: ldr r0{,} [pc{,} \#0]/black,
			~~<2+>: bx r0/black,
			~~<4+>: .word 0x08090055/black,
			~~<8+>: .word 0x08090065/black%
		};

	\def\insnlistpatch{
			\textbf{0x08090030:}/blue,
			~~<0+>: ldr r0{,} [pc{,} \#4]/black,
			~~<2+>: bx r0/black,
			~~\textbf{<4+>: .word 0x08090065}/red,
			~~<8+>: .word 0x08090065/black%
		};
	
	\def\remaplist{
			0x00000000/black,
			0x00000000/black,
			$\cdots$/black,
			\textbf{0x08090065}/red%
		};
	
	\def\reglist{
			\texttt{FP\_CTRL}/\textbf{enabled}/red,
			\texttt{FP\_REMAP}/\textbf{\underline{0x20000000}}/red,
			\texttt{FP\_COMP0}/0x00000000/black,
			$\cdots$/~/black,
			\texttt{FP\_COMP}\textit{n}/\textbf{\underline{0x08090034}}/red%
		};

	\draw[color=black, fill=gray!20, thin] (1.75, -4.785) rectangle (5.75, -6.105);
	\draw[color=black, fill=gray!20, thin] (1.75, -.495) rectangle (5.75, -1.815);

	\draw[color=black, fill=gray!20, thin] (0, -2.495) rectangle (3.5, -4.145);
	\draw[color=black, fill=gray!20, thin] (5, -2.495) rectangle (7.5, -3.815);

	\draw[color=black, fill=white!10, thin] (1.75, -4.785)
			-- ++(0, .33)
			-- ++(2.1, 0)
			-- ++(0.05, -.33);
	\draw[color=black, fill=white!10, thin] (1.75, -.495)
			-- ++(0, .33)
			-- ++(2.1, 0)
			-- ++(0.05, -.33);
	\draw[color=black, fill=white!10, thin] (0, -2.495)
			-- ++(0, .33)
			-- ++(2.1, 0)
			-- ++(0.05, -.33);
	\draw[color=black, fill=white!10, thin] (5, -2.495)
			-- ++(0, .33)
			-- ++(2.1, 0)
			-- ++(0.05, -.33);

	\node[xshift=2pt, inner sep=0pt, color=blue,anchor=west,font=\footnotesize]
			at(0, -2.33) {\texttt{\textbf{FPB Regs}}};
	\node[xshift=2pt, inner sep=0pt, color=blue,anchor=west,font=\footnotesize]
			at(5, -2.33) {\texttt{\textbf{Remap Table}}};

	\foreach \insn\colour [count=\i] in \insnlist  {
		\node[style=datalbl,color=\colour] at(1.75, -0.33*\i) {\insn};
	}

	\foreach \insn\colour [count=\i] in \insnlistpatch  {
		\node[style=datalbl,color=\colour] at(1.75, -4.29 -0.33*\i) {\insn};
	}

	\foreach \insn\colour [count=\i] in \remaplist  {
		\node[style=datalbl,color=\colour,text width=2cm] at(5, -2.33-0.33*\i) {\insn};
	}

	\foreach \reg\val\colour [count=\i] in \reglist {
		\node[style=insnlbl] at(0, -2.33-0.33*\i) {\reg};
		\node[style=datalbl,color=\colour] at(1.5, -2.33-0.33*\i) {\val};
	}

	\draw[-latex'] (1.65, -1.32) -- ++(-2.25, 0)
		node[anchor=south west,inner sep=2pt, text width=2cm, draw, rectangle
		callout, rounded corners=2pt, callout absolute pointer={(0.525, -1.22)},
		fill=white, shift={(0, .5)},dotted, text centered]
			{\footnotesize FPB unit intercepts data fetch at \texttt{0x08090034}}
		|- (-0.1, -2.66-0.33*4);

	\draw (3.6, -2.825) -- ++ (.1, 0) -- (4.8, -2.495) -- ++ (.1, 0);
	\draw (3.6, -3.155) -- ++ (.1, 0) -- (4.8, -3.815) -- ++ (.1, 0);

	\draw[-latex'] (7.6, -3.65) -- ++ (0.5, 0) |-
		node[draw, dotted, rectangle callout, rounded corners=2pt, fill=white, anchor=north east, inner sep=2pt,
		text width=2cm, shift={(0, -.5)},
		callout absolute pointer={(6.975,-5.71)}, text centered]
		{\footnotesize FPB replaces with value on Remap Table} (5.85, -5.61);
	
\end{tikzpicture}

%% file: evaluation/evaluation.tex

In order to evaluate our attacks, we developed a series of proof of concepts in
Cortex-M0/M3/M4/M7/M23/M33 based microcontrollers. These cores are popular
among embedded devices. We utilized microcontrollers from different vendors as
testing platforms even though the CPU cores are licensed from ARM.
In this work, we showcase two attacks that are readily available through the usage of the FPB:
\begin{itemize}
	\itemsep0em
	\item\textbf{MPU Bypass.} We defeat dynamic MPU defenses
	(\S\ref{sec:mpu_bypass}).
	\item\textbf{Kill Protect.} We defeat the protections offered by a
	real-time operating system (\S\ref{sec:kill_protect}).
\end{itemize}


\subsection{Threat Model and Assumptions}\label{sec:threat_model}
\import{threat_model/}{threat_model.tex}

\subsection{Bypassing Memory Protection}\label{sec:mpu_bypass}
A peculiar feature of the FPB is that it ignores any MPU permissions set over
the remap table. This provides an avenue for arbitrary code execution and data
disclosure at various privilege levels. We first describe the case of a
non-privileged application that leaks information from a memory area that can
only be accessed by privileged software. We then describe how we can achieve
arbitrary code execution at a privileged level from a non-privileged program.

We start our experiment by configuring a memory region to be readable only by
privileged software. If we attempt to use a non-privileged load instruction
directly targeting this area, the core generates a \texttt{MemManage}
exception. However, if we configure the FPB's remap table to fall over this area
of memory and then configure a literal comparator to point the address of a
literal used by non-privileged software when this literal is accessed, the FPB
provides a value in the remap table ignoring MPU permissions.
Much like before, though, this mechanism is limited in scope, as remap table
alignment rules apply.

We can also use this FPB behavior to achieve arbitrary code execution and to
leak any privileged region of memory. We set up the FPB to leak the contents of
the vector table as it is always readable regardless of MPU permissions. With
this, we obtain the address of the supervisor call handler. We then set the FPB
instruction comparator register to trap at this address and the corresponding
entry in the remap table to \texttt{0x68004770}. This literal encodes the
instructions \lstinline[style=arma]{ldr r0, [r0]; bx lr}. We point the
corresponding instruction comparator register to the address for the
\texttt{SVCall} handler, replacing the code for \texttt{SVCall} with a load
from anywhere construct. We then set another comparator register to an
arbitrary location in code and the corresponding entry on the remap table to be
\texttt{0xdf004800}, encoding
\lstinline[style=arma]{ldr r0, [pc, #0]; svc #0}. Lastly, using a literal
comparator we set the immediate address to the address in memory we wish to
leak. Upon execution of the \texttt{svc} instruction, our patched interrupt
vector executes. With the CPU in handler mode, we are able to read any
arbitrary privileged location in the address space.

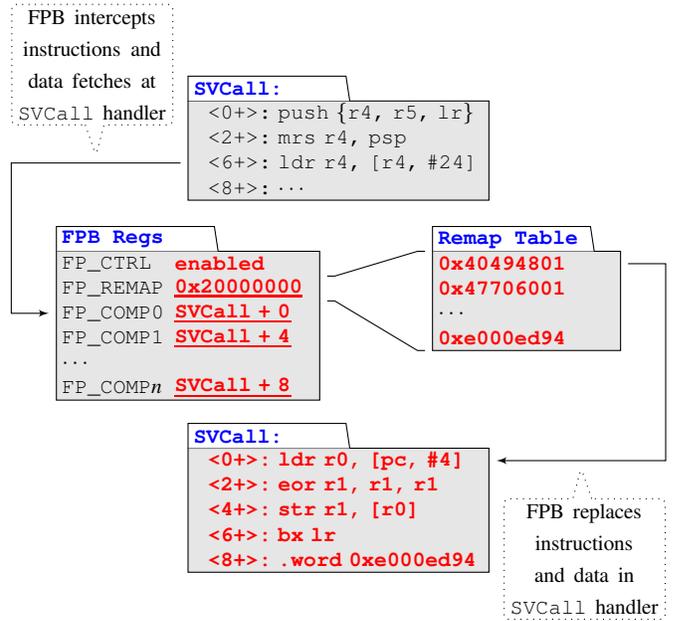
\begin{figure}[h]
    \centering
    \import{./figs/}{fpb_svcall.tex}
	\caption{FPB setup for arbitrary code execution in privileged mode. In this
	example, after our code executes, the MPU is completely disabled on the
	system. For simplicity, we chose to place the data literal at the end of
	the instruction sequence.}
	\label{fig:arbitrary_exec_mpu_disable}
\end{figure}

We extend the same mechanism to achieve arbitrary code execution at privileged
level. Our previous example demonstrates a simple function that leaks data
indirectly. However, using this method we are able to inject code that disables
the MPU. For this, we do a similar setup to the arbitrary information leakage
by overriding the function handler for supervisor calls. However, this time we
utilize two instruction comparator registers and set them as shown in Figure
\ref{fig:arbitrary_exec_mpu_disable}.
We then use a FPB literal comparator register and point it to three words after
the first instruction. We make its entry in the remap table point to the MPU
control register.
The resulting code in the interrupt handler fully disables the MPU in the
system, allowing us to perform code injection and data leakage.

The MPU becomes our indirect target as we wish to account for systems that
dynamically reconfigure it to emulate task isolation. For further testing, we
utilized a version of Minion as presented in \cite{kim2018securing} with a
vulnerability on the real-time operating system core that allows us to set up
the FPB registers. Minion will change the MPU configuration on every task
switch in order to provide isolation. This would require us to disable the MPU
after every task switch. We employ the FPB to override the routines that set up
the MPU regions to disable it instead, achieving persistence across task
switches.

\subsection{Tampering with Real-Time OSes}\label{sec:kill_protect}
\import{./attack/}{rtos.tex}

%% file: threat_model/threat_model.tex
We assume a microcontroller that utilizes a Cortex-M core with FPB
functionality. Our attacks require access to the memory mapped FPB
registers, which are located in the system peripheral bus, requiring privileged
word accesses. We do not require physical access to the platform. An attacker
may gain access and modify the FPB registers using a memory vulnerability such
as the ones described in \cite{CVE-2018-16528, CVE-2018-16526, CVE-2018-16522,
ronen2017iot, cui2013firmware, fur2016exploiting, CVE-2015-6639, CVE-2015-6638,
CVE-2015-4422, CVE-2015-4421,rosenberg2014reflections}
or through fault injection, but neither mechanism is strictly necessary.


Our goal is to bypass any and all security policies deployed in the device,
such as those described in \cite{clements2017protecting, kim2018securing}
without physically altering program code.
We will show how novel use of the FPB recaptures the ability to arbitrarily
leak data
in ways undetectable to the policies at hand.

%% file: figs/fpb_svcall.tex
\begin{tikzpicture}
	\tikzstyle{insnlbl} = [
			text width=4cm,
			inner sep=0pt,
			xshift=2pt,
			font=\footnotesize,
			anchor=west,
		];
	
	\tikzstyle{datalbl} = [
			style=insnlbl,
			font=\footnotesize\ttfamily,
		];

	\def\insnlist{
			\textbf{SVCall:}/blue,
			~~<0+>: push \{r4{,} r5{,} lr\}/black,
			~~<2+>: mrs r4{,} psp/black,
			~~<6+>: ldr r4{,} [r4{,} \#24]/black,
			~~<8+>: $\cdots$/black%
		};

	\def\insnlistpatch{
			\textbf{SVCall:}/blue,
			~~\textbf{<0+>: ldr r0{,} [pc{,} \#4]}/red,
			~~\textbf{<2+>: eor r1{,} r1{,} r1}/red,
			~~\textbf{<4+>: str r1{,} [r0]}/red,
			~~\textbf{<6+>: bx lr}/red,
			~~\textbf{<8+>: .word 0xe000ed94}/red%
		};
	
	\def\remaplist{
			\textbf{0x40494801}/red,
			\textbf{0x47706001}/red,
			$\cdots$/black,
			\textbf{0xe000ed94}/red%
		};
	
	\def\reglist{
			\texttt{FP\_CTRL}/\textbf{enabled}/red,
			\texttt{FP\_REMAP}/\textbf{\underline{0x20000000}}/red,
			\texttt{FP\_COMP0}/\textbf{\underline{SVCall + 0}}/red,
			\texttt{FP\_COMP1}/\textbf{\underline{SVCall + 4}}/red,
			$\cdots$/~/black,
			\texttt{FP\_COMP}\textit{n}/\textbf{\underline{SVCall + 8}}/red%
		};

	\draw[color=black, fill=gray!20, thin] (1.75, -4.785-.33) rectangle (5.75,
	-6.105-.33-.33);
	\draw[color=black, fill=gray!20, thin] (1.75, -.495) rectangle (5.75, -1.815);

	\draw[color=black, fill=gray!20, thin] (0, -2.495-.00) rectangle (3.5,
	-4.145-.33);
	\draw[color=black, fill=gray!20, thin] (5, -2.495-.00) rectangle (7.5,
	-3.815-.00);

	\draw[color=black, fill=white!10, thin] (1.75, -4.785-.33)
			-- ++(0, .33)
			-- ++(2.1, 0)
			-- ++(0.05, -.33);
	\draw[color=black, fill=white!10, thin] (1.75, -.495)
			-- ++(0, .33)
			-- ++(2.1, 0)
			-- ++(0.05, -.33);
	\draw[color=black, fill=white!10, thin] (0, -2.495-.00)
			-- ++(0, .33)
			-- ++(2.1, 0)
			-- ++(0.05, -.33);
	\draw[color=black, fill=white!10, thin] (5, -2.495-.00)
			-- ++(0, .33)
			-- ++(2.1, 0)
			-- ++(0.05, -.33);

	\node[xshift=2pt, inner sep=0pt, color=blue,anchor=west,font=\footnotesize]
			at(0, -2.33-0.00) {\texttt{\textbf{FPB Regs}}};
	\node[xshift=2pt, inner sep=0pt, color=blue,anchor=west,font=\footnotesize]
			at(5, -2.33-0.00) {\texttt{\textbf{Remap Table}}};


	\foreach \insn\colour [count=\i] in \insnlist  {
		\node[style=datalbl,color=\colour] at(1.75, -0.33*\i) {\insn};
	}

	\foreach \insn\colour [count=\i] in \insnlistpatch  {
		\node[style=datalbl,color=\colour] at(1.75, -4.29 -0.33*\i-.33) {\insn};
	}

	\foreach \insn\colour [count=\i] in \remaplist  {
		\node[style=datalbl,color=\colour,text width=2cm] at(5,
		-2.33-0.33*\i-.00) {\insn};
	}

	\foreach \reg\val\colour [count=\i] in \reglist {
		\node[style=insnlbl] at(0, -2.33-0.33*\i-.00) {\reg};
		\node[style=datalbl,color=\colour] at(1.5, -2.33-0.33*\i-.00) {\val};
	}

	\draw[-latex'] (1.65, -1.32) -- ++(-2.25, 0)
		node[anchor=south west,inner sep=2pt, text width=2cm, draw, rectangle
		callout, rounded corners=2pt, callout absolute pointer={(0.525, -1.22)},
		fill=white, shift={(0, .5)},dotted, text centered]
			{\footnotesize FPB intercepts instructions and data fetches at
			\texttt{SVCall} handler}
		|- (-0.1, -2.00-0.33*4);

	\draw (3.6, -2.825-.00) -- ++ (.1, 0) -- (4.8, -2.495-.00) -- ++ (.1, 0);
	\draw (3.6, -3.155-.00) -- ++ (.1, 0) -- (4.8, -3.815-.00) -- ++ (.1, 0);

	\draw[-latex'] (7.6, -3.65+.99) -- ++ (0.5, 0) |-
		node[draw, dotted, rectangle callout, rounded corners=2pt, fill=white, anchor=north east, inner sep=2pt,
		text width=2cm, shift={(0, -.5)},
		callout absolute pointer={(6.975,-5.71+.33)}, text centered]
		{\footnotesize FPB replaces instructions and data in
		\texttt{SVCall} handler} (5.85, -5.61+.33);
	
\end{tikzpicture}

%% file: attack/rtos.tex
We further experiment with the separation model provided by ARM mbed uVisor
\cite{mbed-uvisor}. Imperative to the uVisor model is the existence of a
privileged Hardware Abstraction Layer (HAL) that is assumed to be trusted.
For our demonstration, we create two containers using the uVisor model. One of
them holds a buffer with user-provided input, while the other one performs
requests to the vulnerable handler. Neither container performs inter-process
communication while executing.

Our objective is to undermine the protection mechanism provided by uVisor, that
is we wish to disable the MPU and have unfettered access to the entire device.
For this end, we utilize the container that gathers user input to inject the
remap table with the encoding for the instructions that disable the MPU. The
injected instruction stream is similar to the one covered in Figure
\ref{fig:arbitrary_exec_mpu_disable}. To gain access to the FPB, we exploit an
artificially crafted vulnerability in the trusted HAL which is exposed to
applications through the supervisor call interface. This emulates previously
found vulnerabilities in privileged code \cite{ronen2017iot, CVE-2018-16528,
CVE-2018-16526}. We point the \texttt{FP\_REMAP} register to the buffer used in
the user input container and set the comparator registers to the addresses
containing the routines that handle the MPU when switching containers. Because
the FPB ignores MPU permissions, executing from the user input buffer will not
trigger a fault.  We are once again free to arbitrarily inject code and execute
it, as well as read any memory location.

During our testing, we noted that if we enabled the FPB too early, the system
would go on to exhibit odd behavior, sometimes triggering a hard fault, and
sometimes executing seemingly random code. This is because the remap table had
not been fully loaded to memory, resulting in the comparator registers trapping
instructions and then using an undefined instruction stream to patch. Because the
processes did not have any built-in synchronization, our solution to this was
to ensure that the remap table had been fully loaded before triggering the
system call that enabled the FPB.

%% file: related/related.tex
\subsection{Vulnerabilities on Embedded Devices}
Cui et al. demonstrated the ability and effects of remotely modifying the
firmware of an HP printer by exploiting the firmware update procedure in
\cite{cui2013firmware}. Updating uses a specially crafted print job containing
the update image. Because of insufficient checks done on the update image, the
authors were able to write arbitrary code to the printer's flash memory chips.
Attackers can use this to replace the firmware with a malicious one that can
transmit the document being printed to a remote location. Similarly, Ronen et
al discovered a flaw on the update procedure used in the Phillips Hue smart
bulb \cite{ronen2017iot}. The flaw, caused by a vulnerability in the ZigBee
stack provided by Atmel, allowed the authors to demonstrate how a single
compromised device can spread malware to its neighbors.

Vulnerabilities found in the TCP/IP stack of FreeRTOS can be exploited by a
remote attacker to leak information or achieve arbitrary code execution
\cite{CVE-2018-16522, CVE-2018-16526, CVE-2018-16528}. The vulnerabilities in
question were present not only on FreeRTOS's source tree, but Amazon's AWS
FreeRTOS, and WHIS OpenRTOS and SafeRTOS, the latter of which is certified for
use in safety critical systems. These vulnerabilities are part of a much larger
set found in the FreeRTOS TCP/IP stack \cite{FreeRTOS-vuln}.

While our work does not explicitly rely on vulnerabilities being present in a
device to modify the memory mapped FPB registers, this is a possible avenue to
gain the necessary access. Vulnerabilities such as these are often included
within the threat model of the defenses we subvert in this paper.

\subsection{Defenses that Provide Memory Isolation}
ARM mbed uVisor is a self contained software hypervisor which creates secure
compartments that are independent of each other on Cortex-M3 and Cortex-M4
microcontrollers \cite{mbed-uvisor}. It uses a least privileged mode to execute
\emph{containers}, user tasks which have a limited access domain, setting a
protected environment using the ARM MPU isolating security-critical peripherals
as well as its own code and data. Containers declare an Access Control List
(ACL) to indicate selected hardware peripherals and memories to which they wish
to have access to. Interaction with uVisor is done through a series of APIs
designed around supervisor calls. Interrupts are handled by unprivileged
handlers that have been previously registered. ARM mbed uVisor only provides a
set of core components to design secure applications. Developers are
responsible for declaring the ACLs and any required hardware abstraction layer
for peripherals.

Minion \cite{kim2018securing} provides similar protection features to those
offered by uVisor while also ensuring that the scheduling constrains of the
RTOS are not violated. Minion overcomes limitations in the MPU by dynamically
configuring regions on every task switch operation in order to simulate process
separation with the effect of limiting memory corruption to the reduced scope
of a task.

Clements et al. propose EPOXY \cite{clements2017protecting} which utilizes the
characteristics of the ARM MPU to generate a system of privilege overlays which
results in different segments of the memory map having a different set of
permissions. EPOXY configures the MPU using a set of programmer specified maps
to restrict accesses to defined areas. Critical peripherals are gated behind
these protected areas, and privileges are locally escalated whenever these
areas must be accessed. Locations where areas are accessed are found using
static analysis in an LLVM compiler pass. This protects peripherals from rogue
accesses by normal user code.

\subsection{Defenses to Prevent Memory Corruption}
EPOXY also adds an embedded implementation of SafeStack
\cite{chen2013safestack} to mitigate stack buffer overflow vulnerabilities. A
region for the stack is reserved in the lower area of the microcontroller's
SRAM separating stack buffers from critical control-flow information stored in
the stack. Also, EPOXY applies compile-time randomization to the binary,
scrambling the position of functions and variables in program code and data
spaces, respectively. This requires the attacker to finding the position of
gadgets necessary to construct a code-reuse attack. In this work, we show how
deficiencies in EPOXY can be exploited in combination with the FPB to bypass
the offered protections.

Other stack protection mechanisms include StackGuard \cite{cowan1998stackguard,
wagle2003stackguard}, which utilizes a stack canary between the stored return
address and the callee's stack frame.
StackGhost \cite{frantzen2001stackghost} leverages register window spills in
the SPARC architecture to back up the spilled registers into an OS guarded
portion of memory.
These approaches rely on the generation of secrets by a runtime or
operating system, as well as the means to safeguard these secrets. As we
demonstrated in this work, the FPB can be used to bypass these protections.

\subsection{Mitigating Memory Corruption}
Control-Flow Integrity (CFI) is a security mechanism that enforces an
application's intended control-flow graph to prevent code-reuse attacks in
light of a memory corruption vulnerability. Different CFI approaches have been
proposed with varying degrees of effectiveness and performance overhead. HAFIX
\cite{davi2015hafix} and its successors \cite{christoulakis2016hcfi,
sullivan2016strategy} modify an embedded LEON3 SPARC CPU, and Intel Siskiyou
Peak core in the case of HAFIX. These approaches add new instructions and a
dedicated subsystem to dynamically track an instrumented program's
control-flow while keeping state in a secure memory location. Nyman
et al. propose CFI CaRE \cite{nyman2017cfi} as a means of providing an
interrupt-aware CFI policy for ARM microcontrollers without the need for any
hardware modifications. CFI CaRE requires code to be instrumented, replacing
all control-flow instructions with calls to a Branch Monitor. It also leverages
the security extensions in the ARMv8-M architecture as means of storing
control-flow metadata.

In general, instrumentation-based CFI require a method to inform the system of
valid control-flow transitions, either with specially crafted instructions and
labels, or by unconditionally calling a monitor to examine execution state.
These approaches is that the code is assumed to be immutable, a guarantee that
we find can be easily bypassed using the FPB.

%% file: conclusion/conclusion.tex
In this paper, we have demonstrated that the Flash Patch and Breakpoint (FPB)
unit can be exploited to bypass control-flow integrity, randomization, and
memory protection defenses, achieving arbitrary control-flow hijacking and
information leakage. We have further shown the feasibility of the FPB based
exploitation in a practical case study that bypasses copy protection used in a
commercial product. We also discussed the possible software-based solutions,
showing them to be impractical when defending against our attacks. This
highlights both the severity of the attack and the urgency to protect the
on-chip debugging units from adversaries.

We provided attacks targeting a wide array of Cortex-M platforms in this paper.
In the future, we aim to investigate the effects of other debug peripherals and
their effects of security policies in microcontrollers.